\documentclass[aps,prd,superscriptaddress,showpacs,twocolumn,preprintnumbers]{revtex4}
\usepackage{epsfig,dcolumn}
\usepackage{graphicx}
\usepackage{color}

\begin{document}

\newcommand{\dlt}{\bigtriangleup}
\newcommand{\beq}{\begin{equation}}
\newcommand{\eeq}[1]{\label{#1} \end{equation}}
\newcommand{\bea}{\begin{eqnarray}}
\newcommand{\eea}[1]{\label{#1} \end{eqnarray}}

\parskip=0.3cm


\title{Explicit Model Realizing Parton-Hadron Duality}

\author{L\'aszl\'o L. Jenkovszky}
\affiliation{Bogolyubov Institute for Theoretical Physics (BITP),
Ukrainian National Academy of Sciences \\14-b, Metrolohichna str.,
Kiev, 03680, Ukraine}
\affiliation{Wigner Research Centre for Physics, Hungarian Academy of Sciences\\
1525 Budapest, POB 49, Hungary}

\author{Volodymyr K.~Magas}
\affiliation{Departament d'Estructura i Constituents de la
Mat\'eria,\\ Universitat de Barcelona, Diagonal 647, 08028 Barcelona, Spain}

\author{J. Timothy Londergan}
\affiliation{Department  of Physics and Center for Exploration of Energy and Matter, Indiana University, Bloomington, IN, 47408 USA}

\author{Adam P. Szczepaniak}
\affiliation{Department  of Physics and Center for Exploration of Energy and Matter, Indiana University, Bloomington, IN, 47408 USA}

\begin{abstract}
We present a model that realizes both resonance-Regge (Veneziano) and parton-hadron
(Bloom-Gilman) duality. We first review the features of the Veneziano model and we
discuss how parton-hadron duality appears in the Bloom-Gilman model. Then we review
limitations of the Veneziano model, namely that the zero-width resonances in the
Veneziano model violate unitarity and Mandelstam analyticity. We discuss how such
problems are alleviated in models that construct dual amplitudes with Mandelstam
analyticity (so-called DAMA models). We then introduce a modified DAMA model, and
we discuss its properties. We present a pedagogical model for dual amplitudes
and we construct the nucleon structure function $F_2(x,Q^2)$. We explicitly  show
 that the resulting structure function realizes both Veneziano and Bloom-Gilman
duality.
\end{abstract}


\maketitle


\section{Introduction} \label{s1}
\begin{figure}[htb]
\includegraphics[width=.5\textwidth]{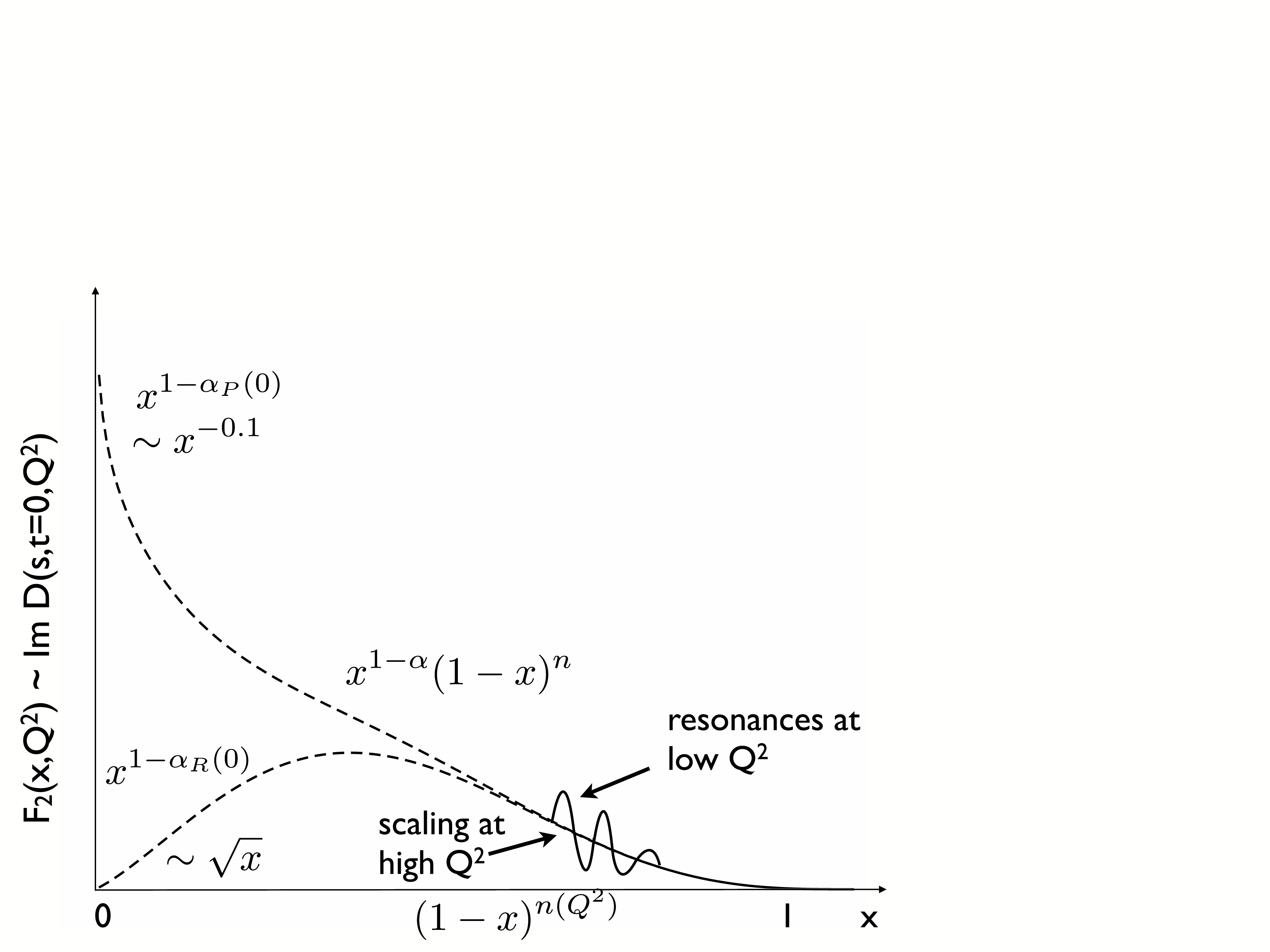}
\includegraphics[width=.5\textwidth]{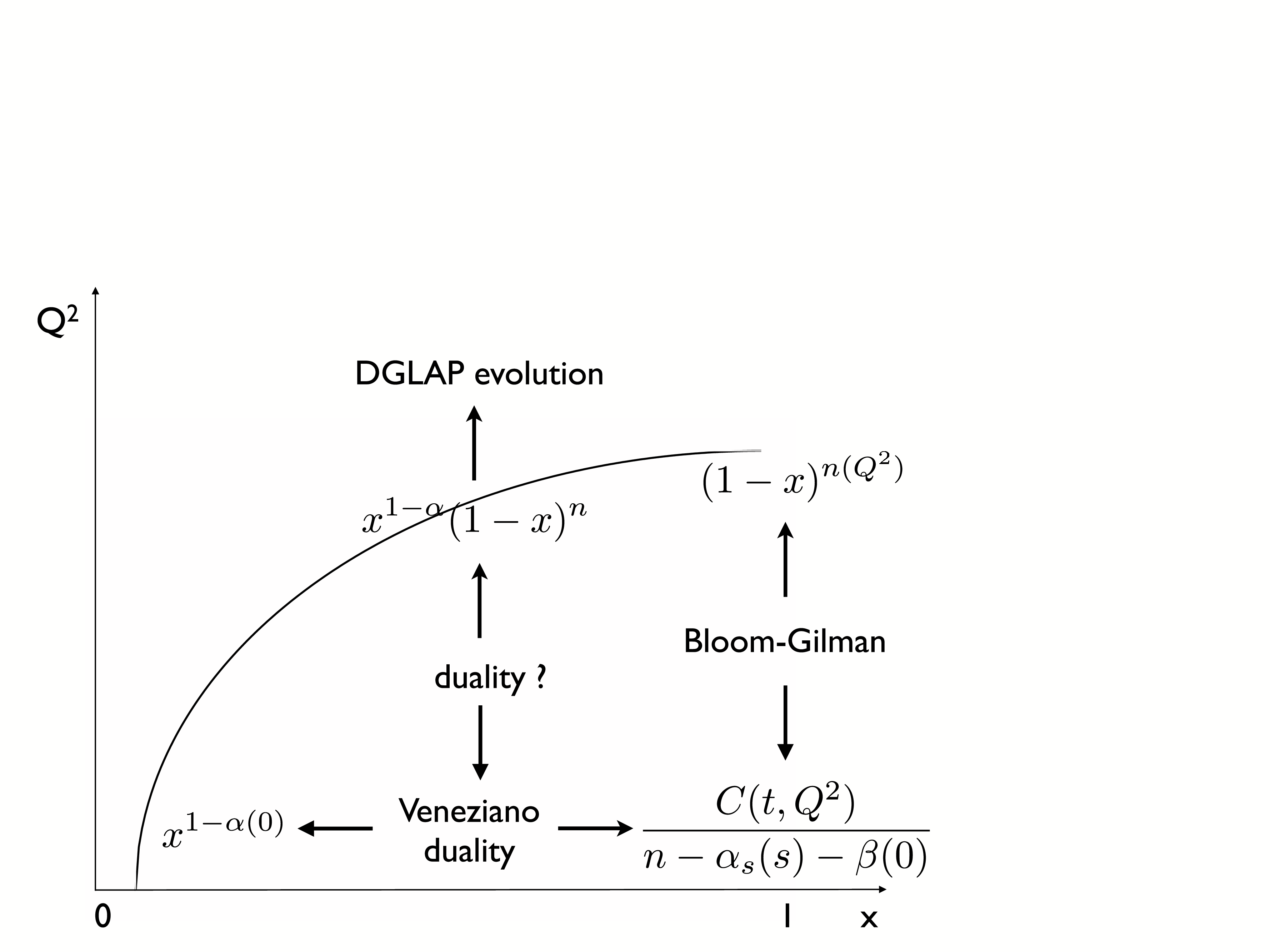}
\caption{Road map visualizing parton-hadron duality for the structure function $F_2(x,Q^2$)
  at  large $x$  (upper panel) and in the whole $x,Q^2$ plane (lower panel). }
  \label{fig:roadmap}
\end{figure}

The subject of this paper was inspired to a large extent by the paper of Bjorken and Kogut
\cite{BK}, which argued that the dynamics of the strong interaction should be continuous
across all regions of energy and momentum transfer. Thus one searches for a unified
description of scattering processes that can explain the properties of reactions over a
wide range of energies and momentum transfers. A qualitative picture of this is shown
in the "road map" depicted in Fig.~\ref{fig:roadmap}. The upper icon shows the behavior
of the structure function $F_2(x,Q^2)$ vs.~Bjorken $x$ variable. At low energies the high-$x$
region is dominated by inelastic resonances, whereas at high energies the resonances
disappear and are replaced by a power-law behavior.

A classic example of a unified description of scattering is the so-called
"Veneziano duality" \cite{Veneziano}. The Veneziano model, reviewed in
Sec~\ref{sec-Veneziano}, can be expanded in terms of a series of narrow resonances, and
at high energies this amplitude demonstrates Regge behavior. This is shown schematically
in the lower panel of Fig.~\ref{fig:roadmap}. The Veneziano model has
subsequently been extended to incorporate broad resonances and Mandelstam analyticity
\cite{DAMA,Jenk}.

Another interesting and important observation is the duality observed in deep inelastic
scattering (DIS).  At low energies and low virtualities these reactions are described in terms
of hadronic properties, while at high virtuality such processes have a partonic description.
At low energies the reactions are characterized by excitation of nucleon resonances and
are described in terms of hadronic excitations. At high energies a partonic description
is more relevant and one sees a smooth curve in the scaling region. However, the
smooth high-energy scaling curve essentially reproduces the average of the resonance peaks
seen at low energies.    This was first noted by Bloom and Gilman \cite{BG} by analyzing data from SLAC, and has since been confirmed by many measurements, with most recent from JLab~\cite{JLab}. The Bloom-Gilman duality region covers the right-hand side in the lower
 panel of Fig.~\ref{fig:roadmap}. This phenomenon has been studied by several groups
\cite{Rujula, Hoyer, Isgur, Melnitchouk, Carlson, Hoyer1, Hoyer2}.
In particular, a relation between the amplitudes measured in exclusive lepto-production
and the quark content of the nucleon was elucidated in Ref. \cite{Adam}. However, for
parton-hadron duality one is still lacking an explicit expression in terms of an amplitude
like that of Veneziano.  In this paper we propose an explicit model that realizes both resonance-Regge ("Veneziano") and parton-hadron ("Bloom-Gilman") duality.

We adopt a two-component model such as the one employed by Harari and Rosner \cite{HR}.
The Harari-Rosner picture is characterized by a smooth background described by a
direct-channel exotic trajectory, which at high
energies is dual to the vacuum exchange amplitude ({\it i.e.,} the Pomeron trajectory) in
the $t$ channel. Separation and identification of the two components is not an easy task,
but fortunately there exist reactions, such as $J/\Psi$ photo- and electro-production, in
which the scattering amplitude (or the structure function) is dominated by
the diffractive component, as discussed in \cite{JPsi}.
In the present paper we will be interested in the non-diffractive (resonance) component of
the dual,  resonance amplitude. Our main result can be summarized by the following
formula which relates the $F_2$ structure function at low-$Q^2$ to the resonance expansion
  of a Veneziano-type amplitude and at high-$Q^2$ to the parton model,

\begin{eqnarray} \label{eq:main}
& \sum_n\frac{[f(Q^2)]^n \mbox{ Im} \alpha_s(s(x,Q^2))}{ [n-\mbox{Re} \alpha_s(s(x,Q^2))]^2
 +\mbox{Im} \alpha_s(s(x,Q^2))^2} & \nonumber \\
& \Uparrow_{Q^2\rightarrow 0} &  \nonumber \\
 & F_2(x,Q^2) &  \nonumber \\
& \Downarrow_{Q^2\rightarrow\infty} & \nonumber \\
 &  (1-x)^{n(Q^2)}. &
\end{eqnarray}
The parton model limit is characterized by the exponent  $n(Q^2)$  and the hadronic limit
 by  a function of the intercepts, $\alpha_s$  of $s$ channel
trajectories.
For the sake of simplicity we ignore spin dependence in this paper. A fully consistent
treatment of the scattering problem would require us to account for the spin dependence.
However, our goal here is to demonstrate qualitatively a proposed new way of constructing
a "two-dimensionally dual" amplitude.

Our paper is organized as follows. In Sec.~\ref{sec-Veneziano}  we review
the narrow-resonance dual Veneziano model. Despite the qualitative successes of this
picture, the zero-width resonances in this model are manifestly non-unitary and violate
analyticity.  In Sec.~\ref{sec-DAMA} we show how these problems can be avoided by incorporating broad resonances  via complex Regge trajectories and Mandelstam analyticity (the so-called DAMA picture). In  Sec.~\ref{sec-MDAMA}, following Ref.~\cite{MDAMA}, the DAMA model is extended off mass  shell to make it applicable to deep inelastic lepton-hadron scattering. Based on this new
dual amplitude, which we call Modified DAMA (or M-DAMA), a structure function $F_2(x,Q^2)$ is calculated in Sec.~\ref{sf}. There we also show that the M-DAMA amplitude, and the related
structure function, explicitly realize parton-hadron duality. Numerical examples and duality
tests are presented in Sect.~\ref{s5}. A summary and discussion of the results can be found
in Sect.~\ref{summary}.

\section{Dual models}
\label{dual}

In this Section we summarize various dual models that attempt to provide a unified
description of scattering phenomena over widely differing regions of energy and
momentum transfer.

\subsection{Narrow-resonance approximation: the Veneziano amplitude}
\label{sec-Veneziano}

Here we briefly review the basic features of resonance-Regge duality, which is
explicitly realized in the Veneziano model \cite{Veneziano}.  The Veneziano amplitude
is given by
$$
V(s,t) = \int_0^1 dz z^{-\alpha(s)}(1-z)^{-\alpha(t)} =
$$
\beq
B(1-\alpha(s),1-\alpha(t)) =
\frac{\Gamma(1-\alpha(s))\Gamma(1-\alpha(t))}{\Gamma(2-\alpha(s)-\alpha(t))} \ .
\eeq{eq11}
The amplitude $V(s,t)$ can be decomposed in a series of resonance poles:
\begin{equation}
\label{eq12}
V(s,t)=\sum_{n=1}^{\infty}\frac{C_n(t)}{n-\alpha(s)}\,,
\end{equation}
 where
\begin{equation}
\label{eq12a}
C_n(t)= {\Gamma(n+\alpha(t)+1) \over n!\ \Gamma(\alpha(t)+1)} \,.
\end{equation}
Using the Stirling formula we can find the asymptotic behavior of $V(s,t)$,
\begin{equation}
 V(s,t)_{|\alpha(s)| \rightarrow\infty} \rightarrow
[-\alpha(s)]^ {\alpha(t)} \Gamma\bigl(1-\alpha(t)\bigr)
\label{eq13}
\end{equation}
 Now, for small $|t|$ the $\Gamma$ function varies slowly compared with the exponential
one. Therefore, taking into account that the Veneziano model requires linear Regge
trajectories, the amplitude $V(x,t)$ displays Regge asymptotic behavior,
\begin{equation}
\label{eq14}
V(s,t)\sim  s^{\alpha(t)}\,.
\end{equation}
The basic properties of the Veneziano model hold only in the narrow-resonance
approximation, from Eq.~(\ref{eq12}) one has an infinite number of zero-width
resonances. At high energies this produces real and linear Regge trajectories.
As a result the original Veneziano model satisfies neither analyticity nor unitarity.
To remedy the problems of an infinite number of narrow resonances,
non-unitarity and an amplitude that lacks an imaginary part, a generalization of the
Veneziano model was proposed, called dual amplitudes with
Mandelstam analyticity or DAMA \cite{DAMA}. In the next section we will discuss
the properties of the DAMA model.

\subsection{Dual amplitude with Mandelstam analyticity (DAMA)}
\label{sec-DAMA}

The so-called dual model with Mandelstam analyticity (DAMA) was proposed as a
generalization of narrow-resonance dual models such as the Veneziano model.
The DAMA model was introduced to avoid the manifest non-unitarity of narrow-resonance
dual models \cite{DAMA}. In contrast to narrow-resonance dual models, DAMA requires non-linear, complex Regge trajectories. The dual properties of DAMA were studied
in Ref. \cite{Jenk}. The DAMA amplitude \cite{DAMA} is given
by,
\begin{equation}
D(s,t)=\int_0^1 {dz \biggl(\frac{z}{g}
\biggr)^{-\alpha_s(s')-1} \biggl(\frac{1-z}{g}\biggr)^{-\alpha_t(t'')-1}},
\label{dama_eq}
\end{equation}
where $\alpha_s(s)$ and $\alpha_t(t)$ are Regge trajectories in the $s$ and $t$ channel
respectively and we introduce the notation $x'=x(1-z), \ \ x''=xz$ (where $x=s,t,u$).
In Eq.~(\ref{dama_eq}), $g>1$ is a parameter.

The introduction of the integration variable accompanying the Mandelstam variable in the power of Eq. (7) enables (moreover, necessitates!)
the use of nonlinear, complex Regge trajectories (impossible in the Veneziano model), required by unitarity and analyticity.
The specific form of these functions, called {\it homotopies}, map the physical trajectory onto a linear function. It is known that unitarity
(and its violation) is particularly sensitive to the singularity structure of the amplitude. DAMA as defined above,
has the pole structure, threshold singularities and the boundary of the double spectral function required by unitarity. For more details see \cite{DAMA}.

In the limit of $s\rightarrow\infty$ and fixed $t$ the DAMA model exhibits Regge behavior,
\begin{equation}
D(s,t)\sim s^{\alpha_t(t)}\,.
\label{dama_regge}
\end{equation}
In the resonance region, the pole structure of DAMA is similar to that of the Veneziano
model except that in the DAMA model multiple poles appear on daughter levels
\cite{DAMA},
\begin{equation}
D(s,t)=\sum_{n=0}^{\infty} D_n(s,t) = \sum_{n=0}^{\infty}
g^{n+1}\sum_{l=0}^{n}\frac{[-s\alpha'_s(s)]^{l}C_{n-l}(t)}
{[n-\alpha_s(s)]^{l+1}}\,,
\label{series}
\end{equation}
where $C_n(t)$ is the
residue, whose form is fixed by the $t$-channel Regge trajectory
(see \cite{DAMA}). The presence of these multiple poles does not contradict the
theoretical postulates.  On the other hand, they can be removed
without any harm to the dual model by means of the so-called Van der
Corput neutralizer \cite{Corput}, corresponding to a slight modification of the integrand in Eq.~(\ref{dama_eq})
and resulting in a standard, "Veneziano-like" pole structure \cite{DAMA},
\begin{equation}\label{eq23-0}
D(s,t)=\sum_n g^{n+1} {C_n(t)\over{n-\alpha_s(s)}}\,.
\end{equation}

The pole term in the DAMA model is a generalization of the Breit-Wigner formula,
and is equivalent to a sequence of resonances lying on a complex trajectory $\alpha_s(s)$.
Such a "Reggeized" Breit-Wigner formula has little practical use in the case of linear
trajectories, since it results in an infinite sequence of poles. However, it becomes a
powerful tool if complex trajectories are used with a bounded  real part and
hence a restricted number of resonances. Moreover, it appears that a small number of
resonances are sufficient to saturate the direct channel.

In contrast to the Veneziano model, the DAMA amplitude presented in Eq.~(\ref{dama_eq}) not
only allows the use of nonlinear complex trajectories, but actually requires the presence
of such trajectories. More specifically, the asymptotic rise of the
trajectories in DAMA is limited by the condition:
\begin{equation}\label{lim}
|{\alpha_s(s)\over{\sqrt s\ln s}}|\leq const, \ \
s\rightarrow\infty.
\label{eq:asympt}
\end{equation}
 The condition in Eq.~(\ref{eq:asympt}) is in accordance with the Froissart
bound, an important upper bound on the rate of rise of cross sections at high energy
\cite{Frois}. Actually, the upper bound in Eq.~(\ref{eq:asympt}) can be
lowered up to a logarithm by requiring wide angle scaling
behavior for the amplitude.

The boundedness of the real part of the trajectories and, consequently, the termination of resonances lying on it,
differs from the general prejudice of their indefinite rise. The latter is supported by several sources, namely: a) the Veneziano amplitude and the string model; b) the seemingly linear behavior  of the spectra of low-lying resonances; c) simplicity. On the other hand, the finite widths of resonances require complex trajectories (as in DAMA). Fits to the masses and decay widths of resonances lying on non-nonlinear trajectories can be found in Ref. \cite{a} (for mesons) and \cite{b} (for baryons). For their physical interpretation in terms of (de)confining quarks in a string model see, e.g., \cite{c,d} and references therein.

Having introduced DAMA, an important question is to examine the dual properties
of the model that incorporates broad resonances, which is a radical modification with respect to the original narrow resonance Veneziano model.
This question was studied in detail in Ref. \cite{Jenk}, where it was shown that DAMA
is dual in the sense that a sum of direct channel resonances reproduces Regge behavior and there is no "double counting" typical of interference models. Moreover, as shown in Ref. \cite{Jenk}, a single pole term of DAMA alone may generate Regge behavior.

\subsection{A modified DAMA model}
\label{sec-MDAMA}

\begin{figure}[htb]
\includegraphics[width=.3\textwidth]{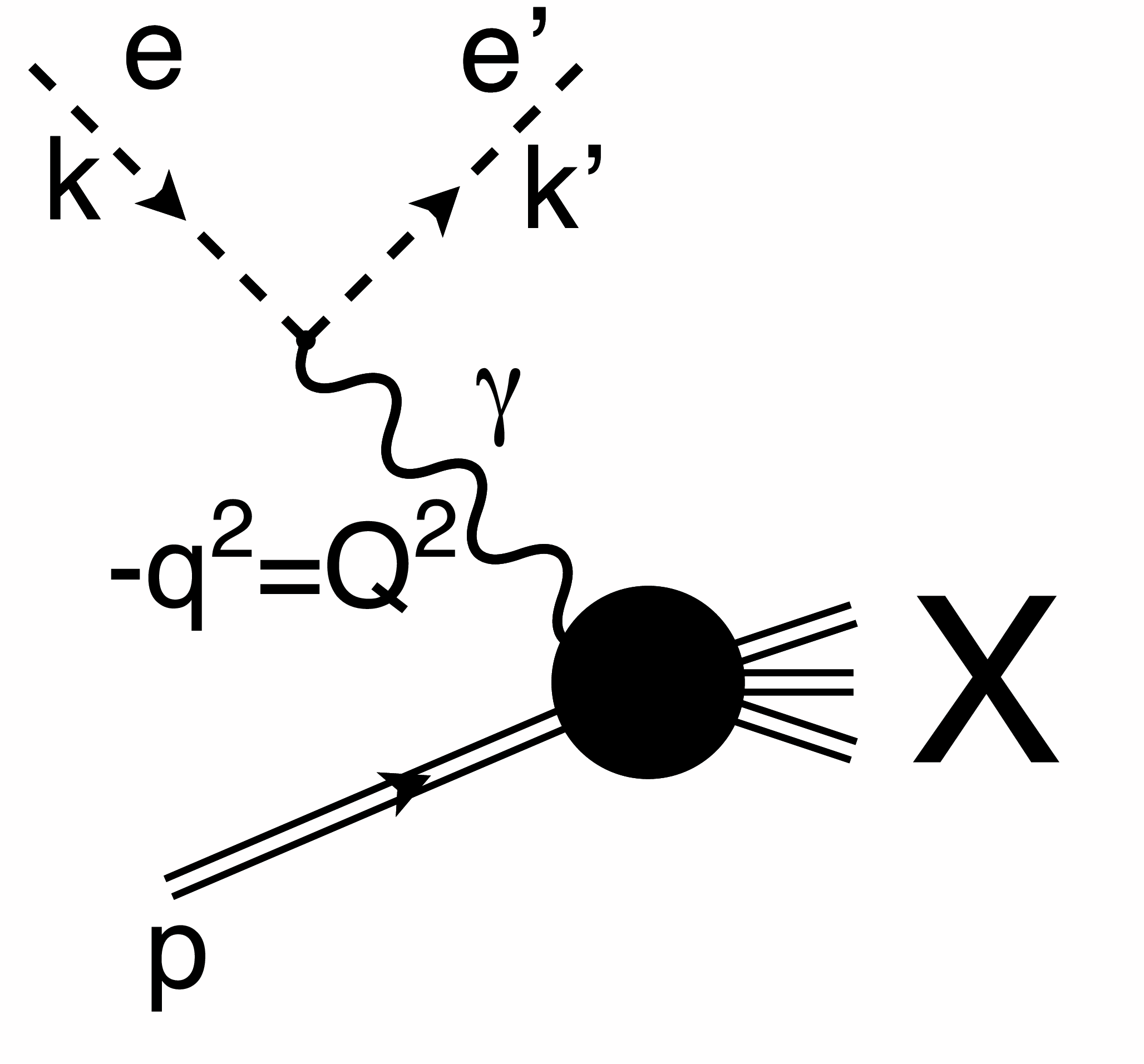}
\caption{Kinematics of inclusive lepton-nucleon scattering. }
\label{rat}
\end{figure}

Inclusive $e^-p$ reactions are most efficiently described in terms of  the following
kinematical variables, the virtuality $Q^2$, $Q^2=-q^2=-(k-k')^2 \ge 0$ of the exchanged photon
  and the Bjorken variable $x= Q^2/(2p\cdot q)$ ({\it cf.} Fig.~\ref{rat}).
  Here $k$ and $k'$ are respectively the initial and final  lepton momentum and $p$ is the nucleon momentum. The variables $x$,  $Q^2$ and the
Mandelstam variable $s$ (of the $\gamma^*p$ system) $s=(p+q)^2$, are not
independent as they obey the relation:
\begin{equation}
s(x,Q^2)=Q^2(1-x)/x+m^2\,,
\label{eq2}
\end{equation}
where $m$ is the proton mass.
In a recent series of papers
\cite{JM0dama,JM1dama,JMfitnospin,JMfitspin} attempts were made to
build a $Q^2$-dependent generalization of the dual amplitude $D(s,t) \to D(s,t,Q^2)$.
This amplitude, a function of three variables, should have correct
known limits, {\it i.e.} it should depend on the on-shell hadronic scattering
amplitude when $Q^2 \sim 0$ on the one hand, and on the nuclear structure
function (SF) when $t=0$, on the other hand. In such a way we hope to complete a unified
"two-dimensionally dual" picture of the strong interactions
\cite{JM0dama,JM1dama,JMfitnospin,JMfitspin} as shown schematically in
Fig.~\ref{fig:roadmap}. The first attempts to combine resonance (Regge) behavior with
Bjorken scaling were made \cite{DG,BEG,EM} at low energies (large
$x$), with emphasis on the $Q^2$-dependence. The amplitudes were chosen such as to match the
known behavior of form factors, of vector meson dominance (VMD) with the requirement of Bjorken scaling \footnote{
VMD is valid in a wide span of low- and intermediate values of
$Q^2)$ (photon virtuality); at very large $Q^2$ deviation from VMD, calculable in perturbative QCD, may be manifest
(for a recent review on the status
of VMD see, e.g. \cite{e}; for a recent update of the HERA results see, e.g., \cite{f}.} .
In the high-energy (low $x$)
region, such a behavior is supported by HERA data. They are discussed in Sect.~\ref{sf}.

While inclusive scattering is determined by the imaginary part in the forward limit, the complete description
requires also the knowledge of $t$ dependence.
In Ref. \cite{JM0dama,JM1dama} the authors attempted to introduce the $Q^2$-dependence into
the Veneziano amplitude \cite{Veneziano} and into the more ambitious Dual Amplitude
with Mandelstam Analyticity (DAMA) model \cite{DAMA}. There were some attempts to introduce  $Q^2$-dependence in the DAMA model either through a $Q^2$-dependent Regge trajectory \cite{JM0dama},
which leads to a problem with its physical interpretation, or through the parameter $g$ \cite{JM0dama,JM1dama}. The latter~\cite{JM1dama} seems more attractive, although it is also restricted by DAMA's intrinsic constraint
$g>1$ \cite{DAMA}. In a series of papers \cite{JM0dama,JM1dama,JMfitnospin,JMfitspin},
the imaginary part of the forward amplitude was related to the total cross section, and to
the nucleon SF.
In this way, the low-$x$ behavior  of the structure function $F_2$ gives a transcendental equation for $g(Q^2)$  (see \cite{JM1dama} for more details). This results in the constraint
$g(Q^2\rightarrow\infty)\rightarrow 0$, inconsistent with the DAMA's constraint $g>1$. Thus, this procedure at best is valid only for a
limited range of $Q^2$ \cite{JM1dama}. An alternative approach is to build a Regge-dual model with $Q^2$-dependent form
factors~\cite{JMfitnospin,JMfitspin}. This procedure was inspired by the pole series
expansion of DAMA that fits the SF data in the resonance region. The goal was to have a
  a $Q^2$-dependent dual amplitude that would lead to such an expansion. The obtained fit was in fairly good agreement with the experimental data, however the attempt to find a
general expression for a two-dimensionally dual amplitude $D(s,t,Q^2)$ failed.

Finally, in Ref.~\cite{MDAMA} a new modified DAMA model with $Q^2$-dependence, thereafter referred to as M-DAMA, was proposed. After the DAMA procedure, this constitutes the 
next step in generalizing the Veneziano model. The M-DAMA model preserves the
attractive features of DAMA, such as its pole structure in $s$
and $t$ and Regge asymptotic behavior. An added feature is that its $Q^2$-dependent form factors have the correct $Q^2\rightarrow \infty$ limit when compared with the structure function (at $t=0$)
 at large-$x$.  The resulting integral representation for M-DAMA  is given by,
\begin{eqnarray}
D(s,t,Q^2)  & = &   \int_0^1 dz \biggl(\frac{z}{g}\biggr)^{-\alpha_s(s')-\beta({Q^2}'')-1} \nonumber \\
 & \times &  \biggl(\frac{1-z}{g}\biggr)^{-\alpha_t(t'')-\beta({Q^2}')-1}\,,
\label{mdama}
\end{eqnarray}
where $\beta(Q^2)$ is a smooth, dimensionless function of $Q^2$ which will be further characterized
 below.   The on-mass-shell limit $Q^2=0$ leads to a shift of the $s-$ and $t-$channel
trajectories by a constant factor $\beta(0)$. Such a shift can be  absorbed into the definition of the Regge trajectories and in this sense the M-DAMA model reduces to DAMA in this limit.
 In  the general case of interaction with a virtual particle with mass $M$ we would have to
 replace $Q^2$ by $(Q^2+M^2)$  in Eq.~(\ref{mdama}). At this point, all of the machinery developed for DAMA (see for example \cite{DAMA}) can be applied to the above integral.
  In particular it can be shown \cite{MDAMA} that the  $s$-channel pole term in the M-DAMA model
 has the following expression ({\it cf.} Eq.~(\ref{series}))
\begin{equation}
D_{n}(s,t,Q^2)=g^{n+1}\sum_{l=0}^{n}\frac{[\beta'(0)Q^2-s\alpha'_s(s)]^{l}C_{n-l}(t,Q^2)}
{[n-\alpha_s(s)-\beta(0)]^{l+1}}\,.
\label{p7}
\end{equation}
The $Q^2$ dependence of the coefficients $C_{l}$   can be directly associated with the form factors
 and will be further discussed in Sec.~\ref{s5}. The presence of the multiple poles, Eq.~(\ref{p7}), does not contradict the
theoretical postulates.  On the other hand, they can be removed without any harm to the
dual model by using the Van der Corput neutralizer, as was discussed in
Sect.~\ref{sec-DAMA}. This results in a "Veneziano-like" pole structure:
\begin{equation}
D_{n}(s,t,Q^2)= g^{n+1}\frac{C_n(t,Q^2)}{n-\alpha_s(s)-\beta(0)}\,.
\label{eq23}
\end{equation}
The asymptotic properties of the M-DAMA model are also similar to those of  DAMA.
 In the Regge limit, ($|s|\rightarrow \infty$, $t,Q^2=\mbox{const.}$), the asymptotic behavior of the M-DAMA model is
\begin{equation}
D(s,t,Q^2)\sim s^{\alpha_t(t)+\beta(0)}g^{\beta(Q^2)} \,, \quad |s|\rightarrow -\infty\,.
\label{W0}
\end{equation}
Thus, in this limit the M-DAMA model has the same asymptotic behavior as
DAMA (except for the shift $\beta(0)$).
It is also interesting to consider the regime that does not exist in DAMA, namely the limit
$Q^2 \to \infty$, with constant $s$, $t$. If one assumes that  $\beta(Q^2) \to  - \infty$ for
$Q^2\rightarrow \infty$, which will be discussed in the following section,
then it can be shown that in this limit \cite{MDAMA}
\begin{equation}
D(s,t,Q^2) \sim  (2g)^{2\beta(Q^2/2)+\alpha_s(s/2)+\alpha_t(t/2)+2}\,.
\label{q3}
\end{equation}
In the case of deep inelastic scattering (DIS), as we shall see below,
if $s$ and $t$ are fixed and $Q^2 \to \infty$ then
$u=-2Q^2\rightarrow - \infty$, which follows from
the kinematic relation $s+t+u=2m^2-2Q^2$.
So, we need also to study the amplitude $D(u,t,Q^2)$ in this limit.
If $|\alpha(-2Q^2)|$ increases
slower than $|\beta(Q^2)|$ or terminates when $Q^2\rightarrow \infty$,
then the previous result (Eq.~(\ref{q3}) with $s$ replaced by $u=-2Q^2$)
is still valid. We will return to these results in the next
section, to check the proposed form of $\beta(Q^2)$.

\section{Nucleon structure function}
\label{sf}
\begin{figure*}[htb]
\includegraphics[width=0.95\textwidth]{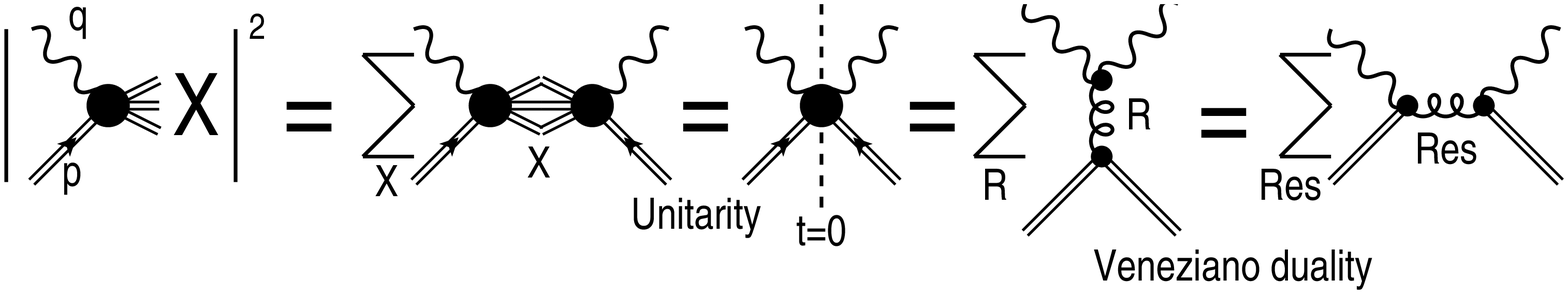}
\caption{According to Veneziano, or resonance-Regge, duality a proper sum
of either $t$-channel or $s$-channel resonance exchanges accounts for
the whole amplitude.}
\label{d2}
\end{figure*}

Having produced a generalization of the DAMA model to describe virtual photon-nucleon 
scattering we discuss its implications for the structure functions.  Figure \ref{d2} 
shows schematically how inelastic lepton-hadron scattering  is related to the forward 
elastic ($t=0$) $\gamma^*p$ amplitude and how the latter can be decomposed into a sum 
of$s-$channel resonance exchanges.

\subsection{ Scaling behavior}

The total cross section for the $\gamma^* p$ reaction is related to the structure function
$F_2(x,Q^2)$ by
\begin{equation}
F_2(x,Q^2)= \frac{Q^2(1-x)}{4\pi \alpha_{em} (1+4m^2 x^2/{Q^2})}
\sigma_t^{\gamma^*p}~,
\label{m23}
\end{equation}
 where
$\alpha_{em}$ is the fine structure constant and we have ignored the
 longitudinal photon cross section which is a reasonable
approximation at high energies.  The total cross section can be related to the imaginary part of the scattering amplitude via the optical theorem,
\begin{equation}
\sigma_t^{\gamma^*p}(x,Q^2)=\frac{8\pi}{p \sqrt{s}} \mbox{Im} A(s,t=0,Q^2),
\label{m22}
\end{equation}
 where $p =  (s-m^2)   \sqrt{1+4m^2 x^2/{Q^2}}/2\sqrt{s}(1-x)$. Thus, we have
\begin{equation}
F_2(x,Q^2)={4Q^2(1-x)^2\over{\alpha_{em} \left(s-m^2\right) (1+\frac{4m^2 x^2}{Q^2})^{3/2}}}
 \mbox{Im}A(s,t=0,Q^2).
\label{f2sigma}
\end{equation}
The minimal model for the scattering amplitude with the proper symmetry properties
is a sum of $s$ and $u$-channel amplitudes, \cite{annalen}
\begin{equation}
A(s,0,Q^2)=(s-u)(D(s,0,Q^2)-D(u,0,Q^2)),
\label{eq34}
\end{equation}
which gives the correct signature in the high-energy limit. Note that $u$ is not an independent variable, since $s+u=2m^2-2Q^2$ or $u=-Q^2(1+x)/x+m^2$. As we remarked earlier, we disregard the  spin and isospin properties of the amplitude in order to concentrate on the dynamics.
In the low-$x$ limit with $t=0$, and $Q^2=\mbox{const}$,
$s=Q^2/x\rightarrow\infty$, $u=-s$, we obtain, with the help of Eqs.~(\ref{W0}),(\ref{eq34}),
\begin{equation}
\mbox{Im} A(s,0,Q^2)| \to s^{\alpha_t(0)+\beta(0)+1}g^{\beta(Q^2)}\,.
\label{eq35a}
\end{equation}
Our philosophy is then as follows, we specify
 $\beta(Q^2)$ in this limit and then use the M-DAMA
 integral representation from Eq.~(\ref{mdama}) to calculate the dual amplitude, and
correspondingly the structure function, in all kinematic domains. We
will see that the resulting SF has qualitatively
the correct behavior in all regions. Moreover, our choice of
$\beta(Q^2)$ will automatically remove the potential posibility of having $Q^2$ poles \cite{MDAMA,Q2poles}.

According to the two-component duality \cite{FH}, both the
scattering amplitude $A$ and the structure function $F_2$ are  sums
of diffractive and non-diffractive terms. At high energies,
both are Regge-behaved. For $\gamma^* p$ scattering only
the positive-signature exchanges are allowed. The dominant contributions come from
the Pomeron and the $f$ Reggeon, respectively. The relevant
scattering amplitude is therefore as follows:
\begin{equation}
A(s,0,Q^2) \propto  \sum_{k=P,R} \xi_k  R_k(Q^2) s^{\alpha_k(0)},
\label{eq4}
\end{equation}
where $\xi_k$, $\alpha_k$ and $R_k$ are respectively the signature,  Regge  trajectory and
residue.  The residue is chosen so as to satisfy approximate Bjorken
scaling for the structure function \cite{BGP,K}. From Eqs.~(\ref{f2sigma}),(\ref{eq4}) the
SF is given by:
\begin{equation}
F_2(x,Q^2)  \propto \sum_{k=P,R} Q^2  R_k(Q^2) s^{\alpha_k(0)-1}
\label{eq4b}
\end{equation}
where $x=Q^2/s$ in the limit $s\to  \infty$. It is obvious from Eq.~(\ref{eq4b}) that Regge asymptotics and scaling behavior require
the residue to fall like $\sim(Q^2)^{-\alpha_k(0)}$. Actually, it could be more involved
if we were to account for scaling violation~\cite{R8,BGP,K}.
Data show that Pomeron exchange leads to a rising structure function at large
$s$ (low $x$). To provide for this we have two options: either to
assume that the Pomeron is supercritical, {\it i.e.} that $\alpha_P(0)>1$ or to assume a
critical ($\alpha_P(0)=1$) dipole or higher multipole Pomeron
\cite{R8,wjs88,Pom1}. The latter condition leads to logarithmic behavior
for the SF \cite{R8,Pom1},
\begin{equation}
F_{2,P}(x,Q^2)\sim Q^2 R_P(Q^2)\ln \Bigl({s\over{m^2}}\Bigr).
\label{eq5b}
\end{equation}
Let us now return to the results of the M-DAMA model. Using Eqs.
(\ref{f2sigma}),(\ref{eq35a}) we obtain,
\begin{equation}
F_2\sim s^{\alpha_t(0)+\beta(0)} Q^2 g^{\beta(Q^2)}\,.
\label{n1}
\end{equation}
Choosing $\beta(0)=-1$, we restore the asymptotic condition of Eq.~(\ref{eq4b}), allowing
us to use the trajectories in their usual form. It is therefore
important to find such a $\beta(Q^2)$ that can provide for
Bjorken scaling.   A possible choice for $\beta(Q^2)$
which smoothly connects the large-$Q^2$ behavior with the low-$Q^2$ data  and has the correct photo-production limit, $\beta(0) = -1$
is given by
\begin{eqnarray}
\beta(Q^2) & = & \beta(0)-\gamma \ln \left(\frac{Q^2+Q_0^2}{Q_0^2}\right) \nonumber \\
 & = &
-1-\frac{\alpha_t(0)}{\ln g} \ln \left(\frac{Q^2+Q_0^2}{Q_0^2}\right)\,.
\label{n3}
\end{eqnarray}
This choice leads to
\begin{equation}
F_2(x,Q^2)\sim
x^{1-\alpha(0)}\Bigl({Q^2\over{Q^2+Q_0^2}}\Bigr)^{\alpha_t(0)}\,,
\label{eq41}
\end{equation}
where the slowly varying factor
$(Q^2/(Q^2+Q_0^2))^{\alpha(0)}$ is typical of
Bjorken scaling violations (see for example \cite{K}).

Next we consider the large-$x$ limit. As $x\to 1$ with
  $s$ fixed, $Q^2=(s-m^2)/(1-x) \rightarrow \infty$ and
correspondingly $u=-2Q^2 \to -\infty$. Using Eqs.~(\ref{q3}),(\ref{f2sigma}),(\ref{eq34}) we find
\begin{eqnarray}
F_2(x,Q^2)  &\propto &  (1-x)^2Q^4g^{2\beta(Q^2/2)} \nonumber \\
& \times & \left(g^{\alpha_s(s/2)}-g^{\alpha_u(-Q^2)}\right)\,.
\label{n4}
\end{eqnarray}
In the limit $Q^2\rightarrow \infty$, the factors
 $g^{\alpha_s(s/2)}$ and $g^{\alpha_u(-Q^2)}$ are slowly varying
functions of $Q^2$ under our assumption about $\alpha_u(-Q^2)$.
Thus, at large $x$ and large $Q^2$ the structure function has the power-law behavior
\begin{equation}
F_2(x,Q^2) \propto  \left(\frac{2Q_0^2}{Q^2}\right)^{2\gamma \ln 2g} \sim
(1-x)^{2\alpha_t(0)\ln 2g/\ln g}.
\label{n5}
\end{equation}

\subsection{ Resonances in the Structure Function}
\label{s5}

We next consider the resonance region structure function $F_2$ in the M-DAMA model.
The  appearance of resonances in the SF at large-$x$ is not surprising by itself,  it follows from Eqs.~(\ref{m22}) and (\ref{f2sigma}). For simplicity we apply the Van der Corput neutralization procedure that was discussed  in Sect.~\ref{sec-DAMA}. Therefore, the pole terms appear in the form specified in  Eq.~(\ref{eq23}). In the vicinity of the resonance energy $s=s_{R},$ only the resonance
term $D_{R}(s,t=0,Q^2)$ is important in the scattering amplitude and correspondingly in
the SF.
Using  $\beta(Q^2)$ in the form given in Eq.~(\ref{n3}), which produces
Bjorken scaling at large $s$, one can obtain the residues at the resonance positions
(see Ref. \cite{MDAMA} for details):
\begin{eqnarray}
& & C_1(Q^2) =
\left(\frac{gQ_0^2}{Q^2+Q_0^2}\right)^{\alpha_t(0)} \nonumber \\
& & \times  \left[\alpha_t(0)+ \frac{Q^2}{Q^2+Q_0^2} \ln g  - \frac{\alpha_t(0)}{\ln g}
\ln \left(\frac{Q^2+Q_0^2}{Q_0^2}\right)\right]. \nonumber \\
\label{c1}
\end{eqnarray}
The term in front of the bracket provides the typical  $Q^2$-dependence for the form factor, while the remaining part  gives an additional slowly varying dependence on $Q^2$. Higher orders of, $C_n$ for subleading resonances,
have the same primary $Q^2$ dependence: the $(Q_0^2/(Q^2+Q_0^2))^{\alpha(0)}$ form factor.
 By introducing  $Q^2$-dependence through the parameter $g$, we would find, however,
  a significant difference.
  As  can be seen from Eq.~(\ref{eq23}), $g$ enters with different powers for different
resonances on a single trajectory, with the powers increasing in steps of two. Thus, if
$g\sim (Q_0^2/(Q^2+Q_0^2))^{\dlt}$, then the form factor for the first ($n=0$) resonance is
 proportional to $(Q_0^2/(Q^2+Q_0^2))^{\dlt}$, while the second one
 is proportional to  $(Q_0^2/(Q^2+Q_0^2))^{3\dlt}$,  and so on. As discussed in \cite{JMfitspin} the present accuracy of the data unfortunately does not allow one to discriminate between constant powers of the form factor and increasing ones~\cite{Stein,Niculescu,Osipenko,DS}.

Let us give a quantitative example of how resonances appear in the structure function. For this simplified discussion we assume that our dual amplitude is a sum of three
resonances on some $s$-channel trajectory, and we also assume a simplified form of the form factor discussed above. From
Eq.~(\ref{eq23})  we find the amplitude proportional to
\begin{equation}
D(s,t=0,Q^2) \propto  \sum_{n=1}^{3} g^{n+1} \left(\frac{gQ_0^2}{Q^2+Q_0^2}\right)^{\alpha_t(0)}
\frac{1}{n-\alpha_s(s)+1}\,.
\label{D_3res}
\end{equation}
For the Regge trajectory, for simplicity we assume a linear form with a square root 
branch point at the pion-nucleon threshold,  $s_0=(m_\pi+m)^2$,
\begin{equation}\label{test_traj}
\alpha_s(s)= \alpha_s(0) + \alpha'_s(0) s+\gamma(\sqrt{s_0}-\sqrt{s_0-s})\,,
\end{equation}
with typical values $\alpha_s(0) = 0.1$, $\alpha'_s(0) =1 \mbox{ GeV}^{-2}$ and 
$\gamma = 0.1 \mbox{ GeV}^{-1}$ for the intercept and  slope of the trajectory and width 
of the resonances, respectively \footnote{More advanced, realistic examples of 
non-linear trajectories can be found in Refs.\cite{a,b}.}.

\begin{figure}[htb]
\includegraphics[width=0.5\textwidth]{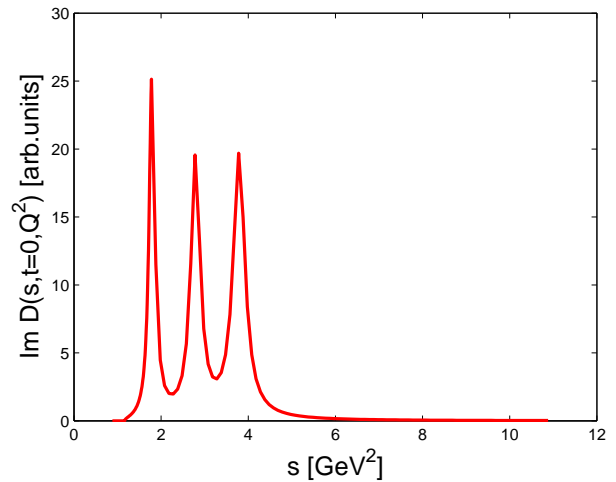}
\caption{(color online) Imaginary part of the dual amplitude (in arbitrary units) of 
Eq.~(\ref{D_3res}) as a function of the Mandelstam variable $s$.}
\label{dual_s}
\end{figure}

\begin{figure}[htb]
\includegraphics[width=0.5\textwidth]{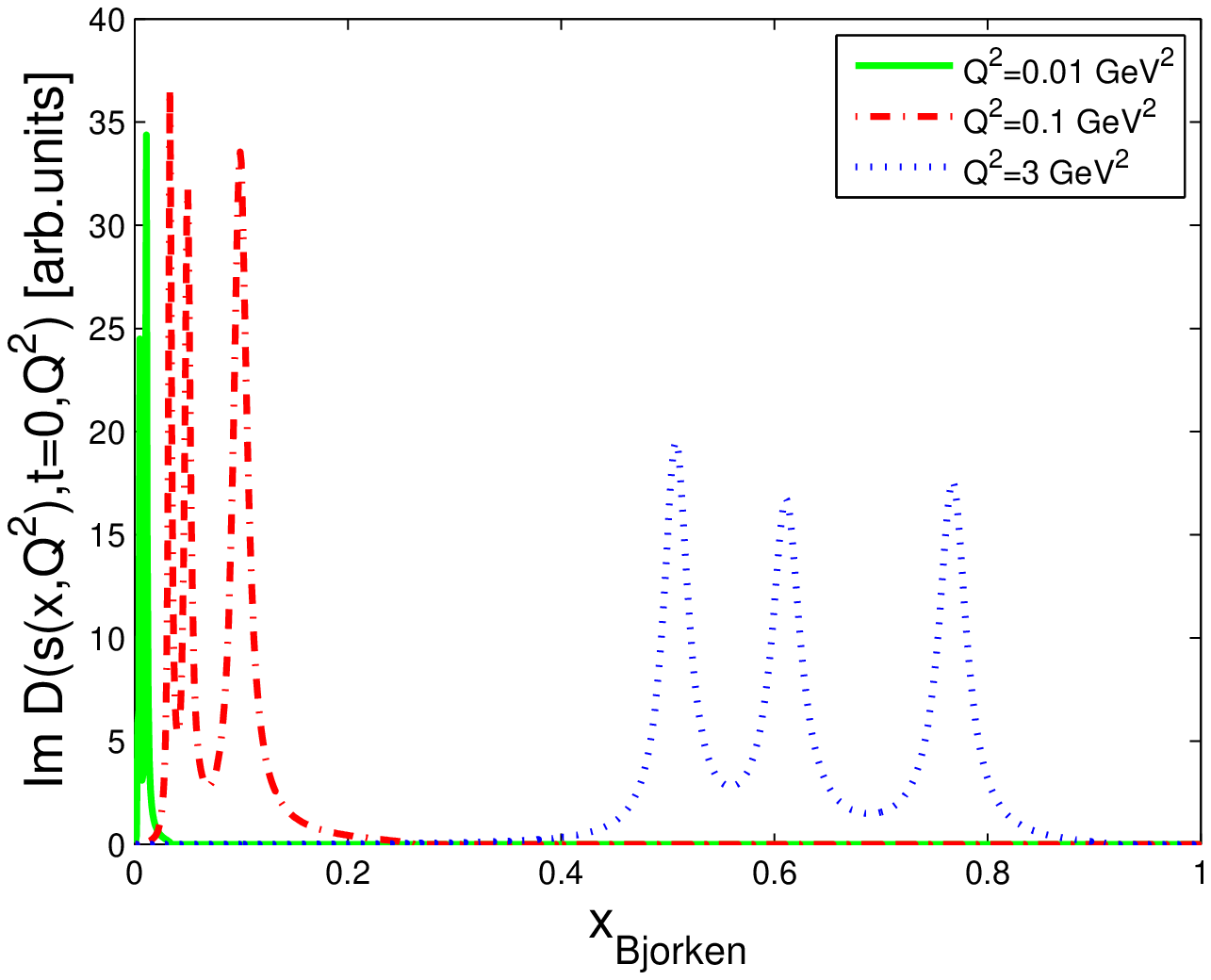}
\includegraphics[width=0.5\textwidth]{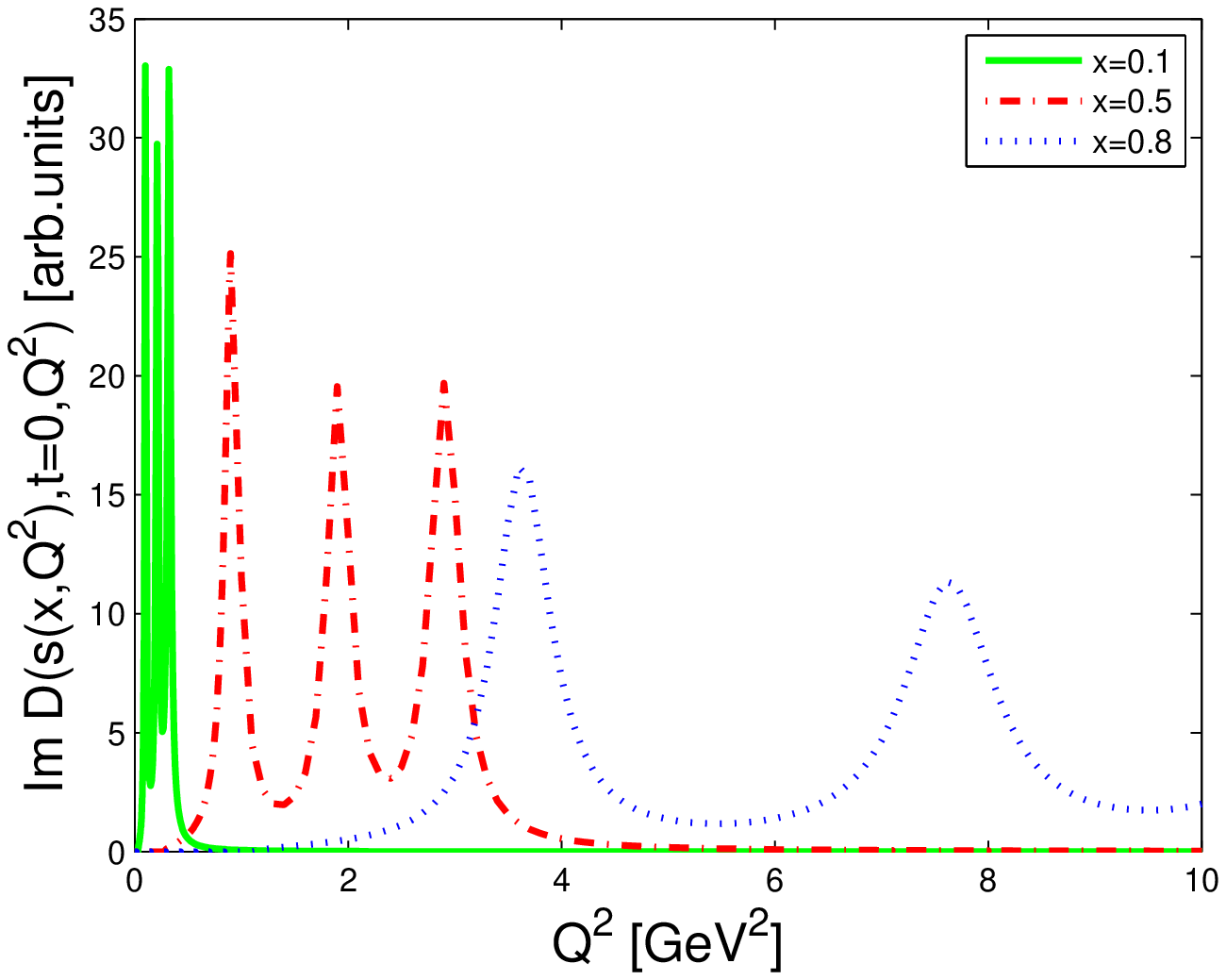}
\caption{(color online) 
Properties of the dual amplitude, given by Eq.~(\ref{D_3res}) as a function of Bjorken
$x$ (top panel) for  various values of $Q^2$ and as a function of the virtuality $Q^2$ 
for various values of $x$ (lower panel).}
\label{dual-x-q2}
\end{figure}

\begin{figure*}[htb]
\includegraphics[width=0.95\textwidth]{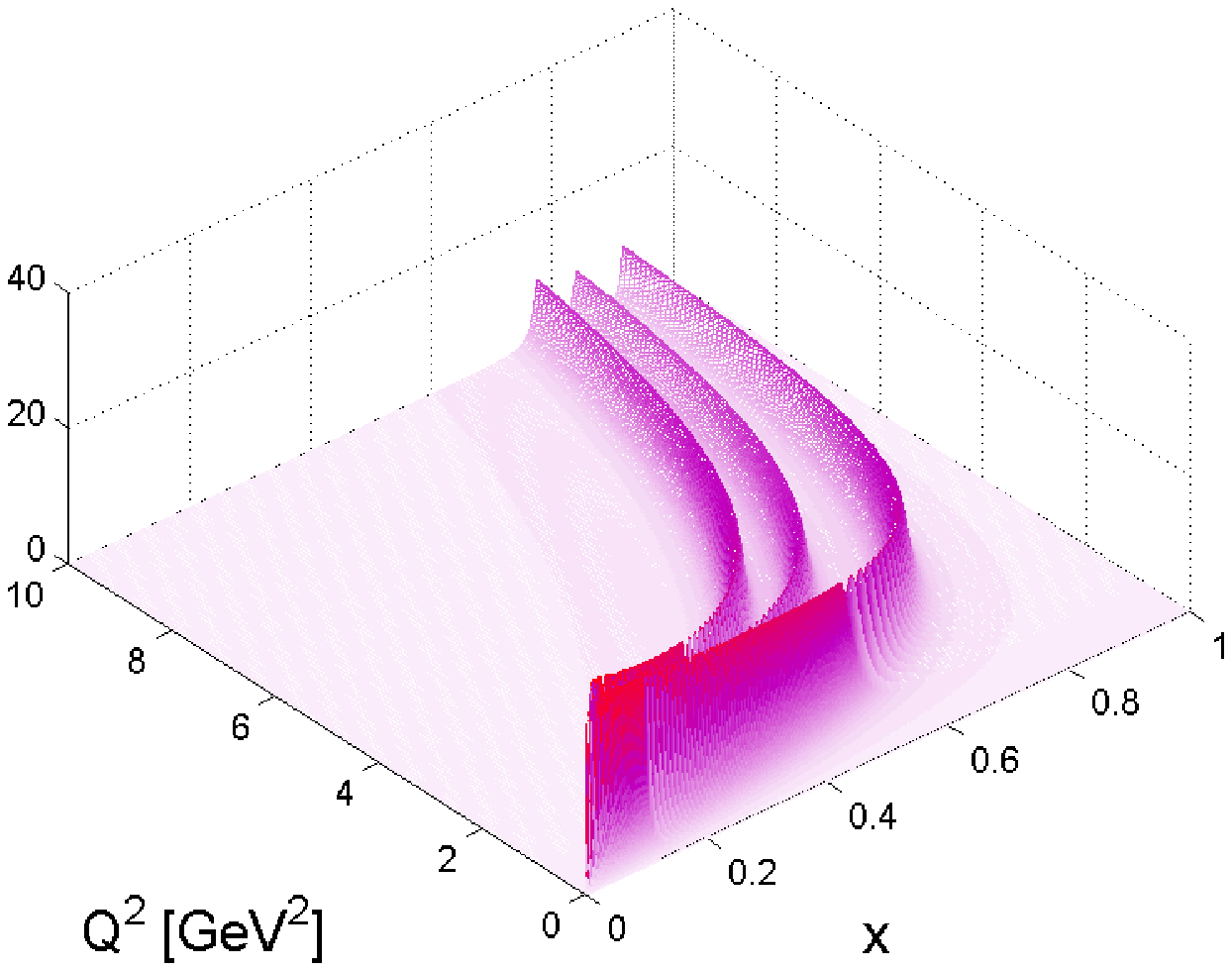}
\caption{(color online) Position of the three resonances in the imaginary part of the 
dual amplitude on the $x-Q^2$ plot.}
\label{3D}
\end{figure*}

\begin{figure*}[htb]
\includegraphics[width=0.95\textwidth]{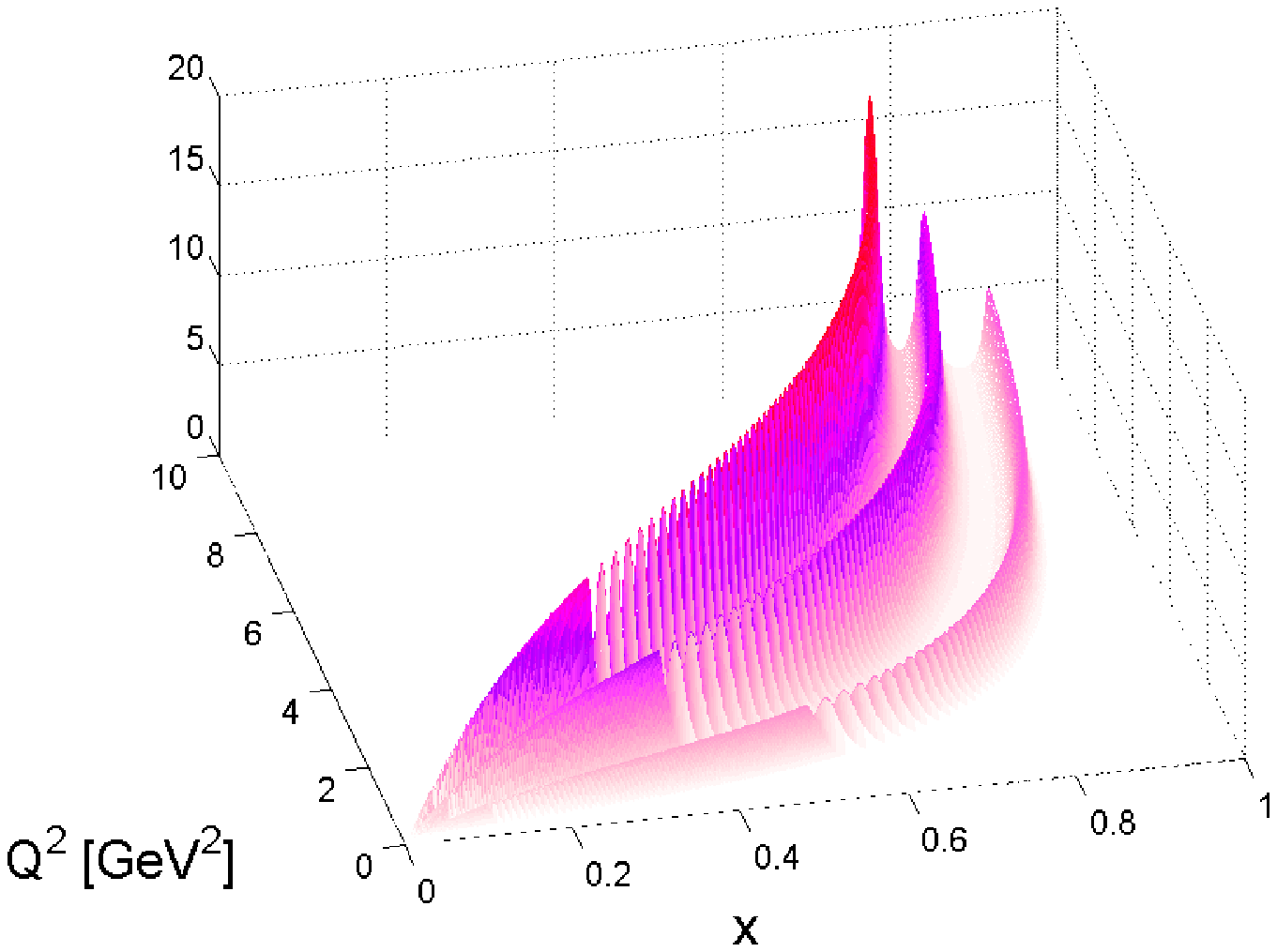}
\caption{(color online) The structure function $F_2$ (in arbitrary units) as a function 
of $Q^2$ and $x$.}
\label{SF_3D}
\end{figure*}

\begin{figure}[htb]
\includegraphics[width=0.5\textwidth]{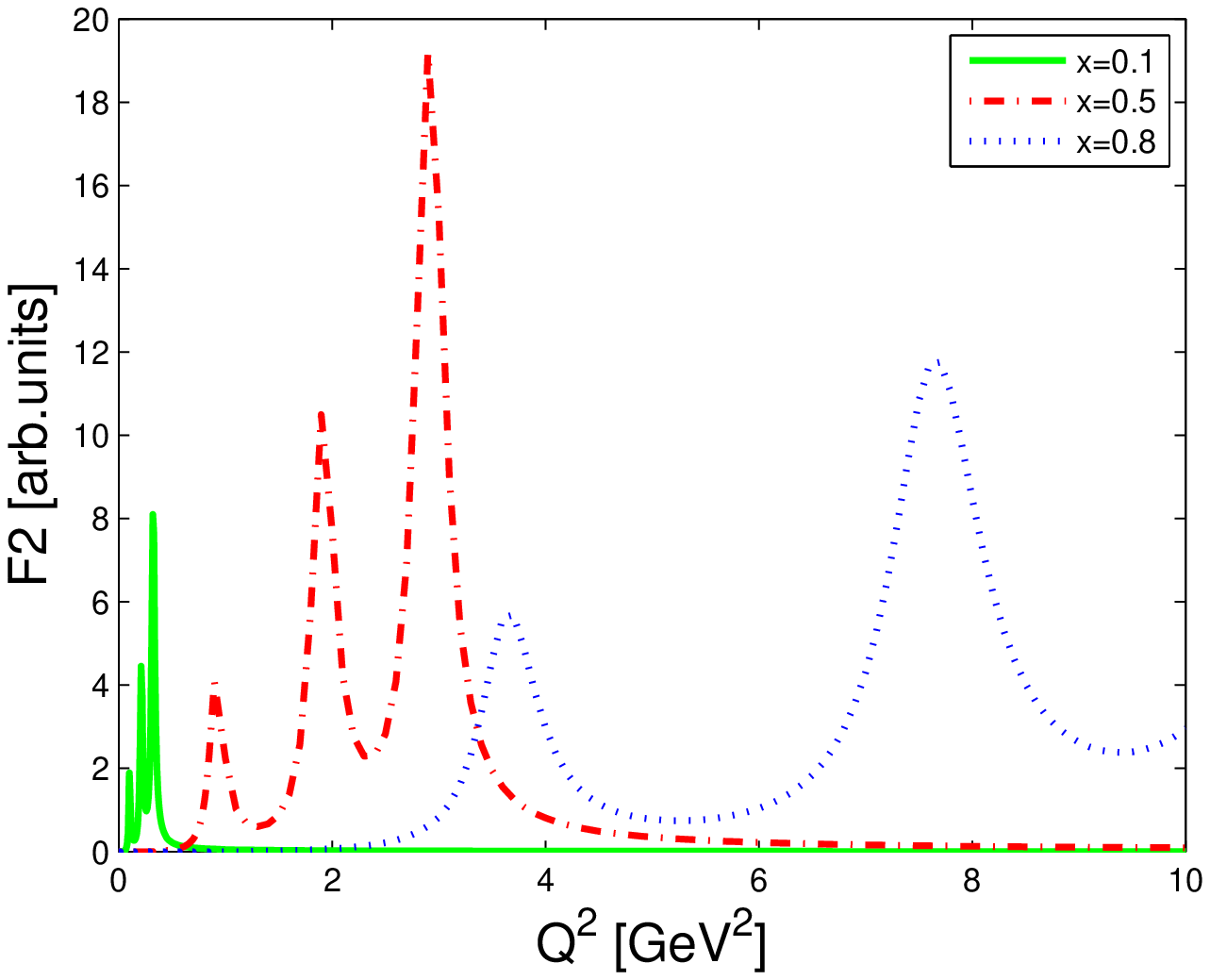}
\includegraphics[width=0.5\textwidth]{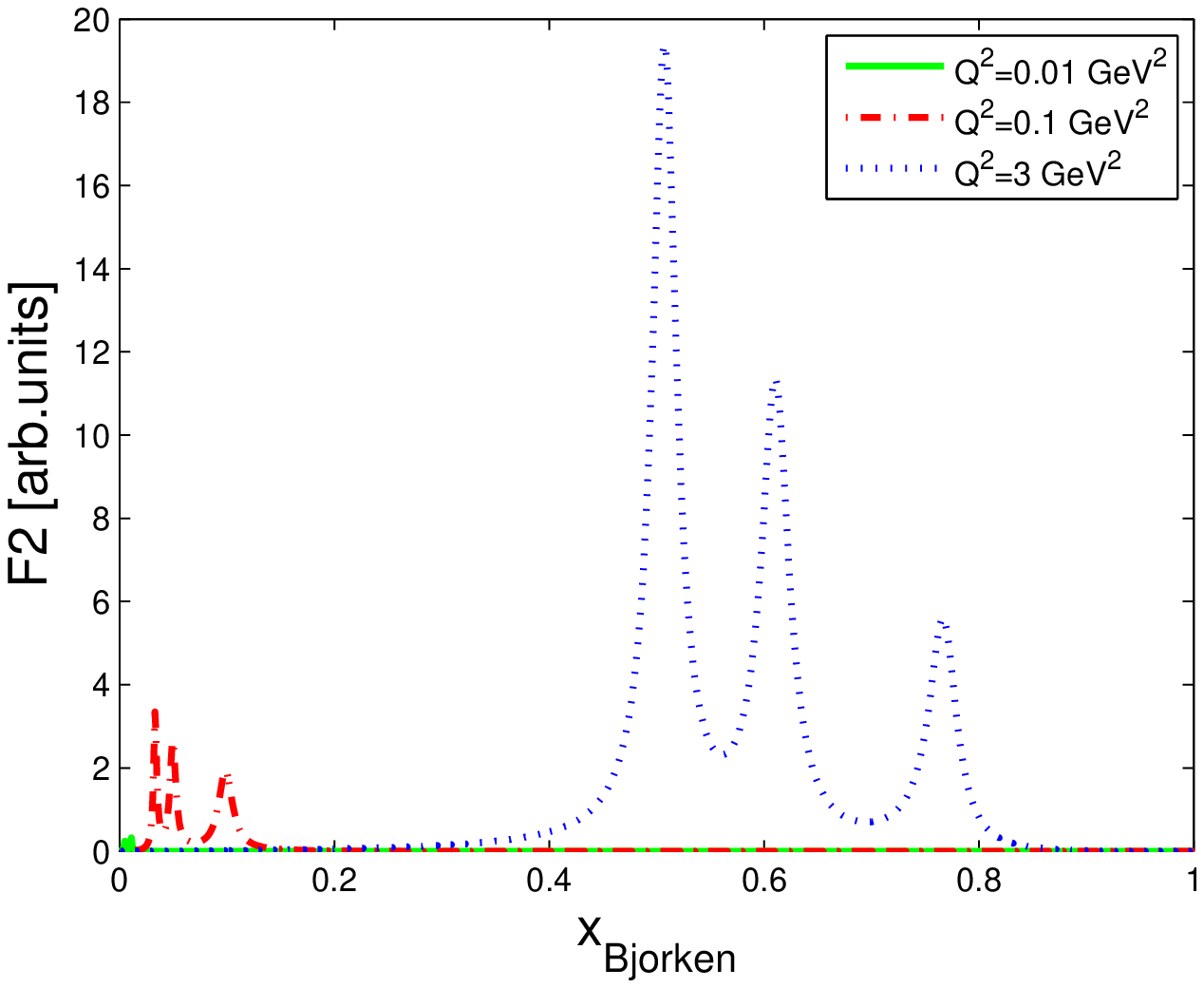}
\caption{(color online) Properties of the structure function (SF) $F_2(x,Q^2)$ of
Fig.~\ref{3D} for various kinematics, as a function of $Q^2$ (top panel) and
 as a function of Bjorken $x$ (lower panel). }
 \label{SF_s}
\end{figure}

%
Other parameters of the model are:  $\alpha_t(0)=0.5$, $Q_0^2=1$ GeV$^2$, and $g=1.5$. This choice of the scale parameter $g$  gives three resonance peaks of similar height in $Im\ D(s,t=0,Q^2)$ as shown in Fig.~\ref{dual_s}.
We show how the positions of the resonances move as a function
of $Q^2$ and $x$ in Fig.~\ref{dual-x-q2} and ~\ref{3D}. In Fig.~\ref{dual-x-q2} we plot the
  imaginary part of the dual amplitude $D(s,t=0,Q^2)$ as a function of either $x$ or $Q^2$.
  The 2-dimensional plot in Fig.~\ref{3D}  clearly shows how the positions
of the resonances move according to ({\it cf.} Eq.~(\ref{eq2})),
\begin{equation}
s_{R,n} =Q^2(1-x)/x+m^2=\mbox{ const.}
\label{eq2Res}
\end{equation}
Similar plots for the structure function $F_2(x,Q^2)$ can be produced.  We use
Eqs.~(\ref{f2sigma}) and (\ref{eq34}), and we neglect the $u,t$ terms in
Eq.~(\ref{eq34}), which is justified in the domain where the $s$-channel resonances
dominate. In Fig.~\ref{SF_3D} we plot the structure function $F_2(x,Q^2)$, as a function of $x$ and $Q^2$ and in Fig.~\ref{SF_s} we plot various cuts on
the SF $F_2(x,Q^2)$.  We see that, with increasing $Q^2$, the resonances
produce a power-law behavior at large $x$.

Some of the qualitative features of the structure function $F_2(x,Q^2)$ are analogous to
those for the dual amplitude shown in Figs.~\ref{3D} and \ref{dual_s}. With increasing
$Q^2$, the resonance peaks move to higher values of $x$ and their spacing in $x$ also increases.
 Also, with increasing $x$ the  resonances move to higher $Q^2$ and their spacing in $Q^2$ increases, as is shown in Fig.~\ref{SF_s}. However, while the three resonance peaks in the dual amplitude
have nearly equal heights, the resonance peaks in the SF $F_2$ have rather different
heights, which is evident by comparing Figs.~\ref{dual_s} and \ref{SF_s}.

\section{Summary}
\label{summary}

Fig.~\ref{fig:roadmap} schematically shows the "correspondence principle", partly 
realized in our model. The dual amplitude should reduceto a sum of resonance poles 
or smooth functions scaling in $x$ in the relevant kinematical regions, and it should smoothly interpolate between these two regions along the vertical line of the lower icon of Fig.~\ref{fig:roadmap}. The upper icon of the same figure shows this effect on the large-$x$ behavior of the structure function.

In this paper we have presented an explicit model that realizes parton-hadron duality.
In the present model, the same amplitude interpolates between low- and high
$x$, and thus realizes "Veneziano duality", or resonance-Regge duality at low $Q^2$.
The M-DAMA model exhibits "Bloom-Gilman" or parton-hadron duality,
interpolating between resonant behavior at low $Q^2$ and power-law behavior at high $Q^2$
in the large-$x$ region. In this way, the M-DAMA model realizes "two dimensional
duality."

The M-DAMA model, with three resonances approximating the entire
resonance regime is a simple pedagogical example.
Nevertheless, the model successfully incorporates two types of duality (two-dimensional duality). Can the path shown schematically in Fig.~\ref{fig:roadmap} be closed by
relating small- and large-$x$ with some (any?) values of $Q^2$? In other
words, given the off-mass-shell dual amplitude in the M-DAMA model that interpolates between
the above regions, one can ask if the large-$x$ structure function "knows about" its
small-$x$ behavior and vice versa? This would require very strong correlations between
the structure function in different regions. We do not have any convincing answer to this
question but it is an interesting conjecture given the success of the M-DAMA model
that we present in this paper.

The notions of duality can be formulated and applied in a rigorous fashion. Note
that one consequence of duality means that a scattering amplitude cannot be decomposed into
a sum of resonances plus an independent smooth asymptotic term, as is the case in
interference models. This criterion was studied in detail in Ref.~\cite{Jenk}, where
the DAMA model was shown to have genuine dual properties. It would be interesting to
see if a similar criterion and proof for parton-hadron duality could be obtained
for the M-DAMA model.

One of the main virtues of the M-DAMA model is its applicability to physical processes over
a wide kinematic region and connections imposed by the duality conditions. Detailed
quantitative applications of this model will require the imposition of various constraints
on the amplitudes. Our model utilizes two-component duality, in which the scattering amplitude is a sum of a diffractive and non-diffractive term. In the first case, the high-energy
behavior is determined by the exchange of a vacuum (Pomeron) trajectory that is dual to the
direct-channel exotic trajectory. The ordinary $s$ and $t$ channel trajectories are connected with resonance spectra. The trajectories, both leading and non-leading, are strongly bounded non-linear complex functions. Examples of practical applications can be found in Refs.\cite{JM1dama,JMfitnospin,JMfitspin,JPsi}.

The model is not unique, although the form of the dual
amplitude is quite constrained as been discussed above, see also \cite{DAMA, MDAMA, Jenk}. The remaining freedom can be further narrowed by
fitting the model to experimental data. Furthermore, one can compare the model to
other approaches to high-energy reactions, such as the quark counting rules (in the hard region
\cite{JMfitspin}), string models e.g. in the large-$x$ region \cite{Kaidalov}, or QCD. In
this way one can hope to reduce some of the freedom in our dual model and to constrain
its free parameters.

\section*{Acknowledgements}
The work of LLJ was supported by the Hungarian Academy of Sciences' Fellowship for Transfrontiery
Hungarian Science. The work of VKM was partly supported by the contracts FIS2008-01661 from
MICINN (Spain), by the Ge\-ne\-ra\-li\-tat de Catalunya contract
2009SGR-1289, and by the European Community-Research
Infrastructure Integrating Activity ``Study of Strongly
Interacting Matter'' (Hadron Physics2, Grant Agreement n. 227431)
under the Seventh Framework Programme of EU. The work of APS is supported in part by
the US Department of Energy under contract DE-FG0287ER40365, and the research of JTL is
supported by the US National Science Foundation under grant PHY-0854805.

\vfill \eject
\end{document}